\documentclass[prl, aps, english, twocolumn, hyperref,floatfix,showpacs]{revtex4}
\usepackage{graphicx}
\usepackage{amssymb, amsmath, amsfonts, color, rotating, multirow, graphicx, bm}
\usepackage[bookmarks=false,pdfstartview=FitH,hyperindex=true, colorlinks, linkcolor=blue, citecolor=blue]{hyperref}

\usepackage{graphicx}

\begin{document}

\title{Anomalous charge pumping in a one-dimensional optical superlattice}

\author{Ran Wei and Erich J. Mueller}

\affiliation{Laboratory of Atomic and Solid State Physics, Cornell University, Ithaca, New York 14853}

\begin{abstract}
We model atomic motion in a sliding superlattice potential to explore topological ``charge pumping"
and to find optimal parameters for experimental observation of this phenomenon. We analytically study the band-structure,
finding how the Wannier states evolve as two sinusoidal lattices are moved relative to one-another, and relate this
evolution to the center of mass motion of an atomic cloud. We pay particular attention to counterintuitive or anomalous regimes,
such as when the atomic motion is opposite to that of the lattice.
\end{abstract}

\date{\today}

\maketitle

\emph{Introduction --} 
Slow periodic changes in a lattice potential can transport charge. For a filled band, the integrated particle current per
cycle in such an adiabatic pump is quantized \cite{Thouless1983}. 
We study a simple but rich example of this phenomenon, namely charge transport in a sliding superlattice,
and draw attention to its counterintuitive properties
such as regimes where the charge moves faster than the potential, or even travels in 
the opposite direction. We argue that this anomalous transport is observable in a cold atom experiment.

The quantum mechanics of particles in a one-dimensional (1D) superlattice is
rich, displaying a fractal energy spectrum \cite{Hofstadter1976}, and for incommensurate periods
boasting a localization transition similar to what is seen in disordered lattices \cite{Andre1980}.
While recent studies have focused on the tight-binding limit (the Aubry-Andre model)
\cite{Shu2012, Shu2013a, Shu2013b, Shu2014,Fleischhauer2013,
Sarma2013, Chong2015, Zilberberg2012,Kraus2012,Zilberberg2013,Goldman2012,
Duan2013,Brouwer2013,Deng2014,Matsuda2014,Gaspard2014,Ortix2014,Naumis2013},
we study the continuous limit of the 1D superlattice,
where, because of the weak potential, the single-particle spectra can be calculated perturbatively.
Related cold atom proposals on quantized transport 
\cite{Ripoll2007,chuanwei2011,Zhu2014,Daixi2013,Zhu2015} have focused on the simplest superlattice 
where one sub-lattice constant is half of the other, and are therefore not in
the anomalous regime which interests us.

The 1D superlattice can be mapped onto the Harper-Hofstadter model \cite{Harper1955,Hofstadter1976}.
The topological numbers (Chern numbers) associated with charge pumping can be mapped onto
quantized Hall conductances \cite{TKNN1982,Kohmoto1985}.
Recent experiments involving artificial gauge fields on 2D optical lattice
have aimed to measure these 2D Chern numbers
\cite{Bloch2011, Bloch2013, Atala2014, Bloch2014, Ketterle2013}. 
There are also related studies based on measurement of Hall drift \cite{Goldman2013},
Bloch oscillations \cite{Cooper2013,Xiongjun2013},
Zak phase \cite{Demler2013,Atala2013,Demler2014}, time-of-flight images \cite{Zhao2011,Ripoll2011,Duan2014},
edge states \cite{Gerbier2012,Spielman2013,Gerbier2013,Reichl2014,Spielman2015,Fallani2015}, or density plateaus \cite{Okel2008,Zhu2008}.

In this letter, we study the charge transport in a
1D sliding superlattice, where the moving
lattice period is an arbitrary rational multiple of the static lattice.
We analytically calculate energy band gaps and the topological
invariants which give the integrated adiabatic current 
per pumping cycle \cite{Thouless1983}.
The fact that this current can be made \emph{arbitrarily large} and/or \emph{opposite}
to the direction of the sliding potential is counterintuitive. 
We present a physical interpretation of this phenomenon in
terms of the quantum tunneling of Wannier functions between minima in the potential.
We propose an experiment to detect this anomalous adiabatic current,
and derive the optimal parameters. Through numerical simulations, we confirm that a \emph{negative} integrated
current and a non-trivial Chern number ${\mathcal C}=-1$ 
is readily measured in an experiment.

\begin{figure}[!htb]
\includegraphics[width=8.5cm]{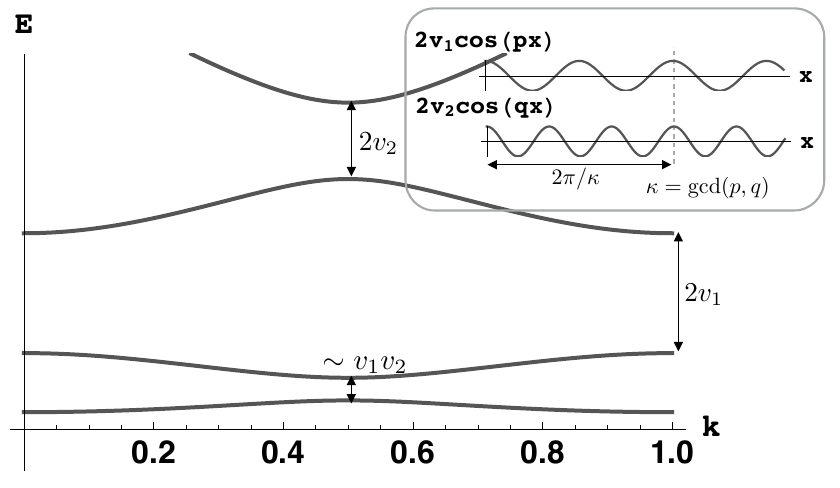}
\caption{Band structure of a 1D superlattice for $p=2, q=3$,
showing energy $E$ vs dimensionless wave-vector $k$ for weak potentials.
Inset shows the two potential making up the superlattice,
and illustrates the unit cell with period set by the greatest common divisor $\kappa\equiv{\rm gcd}(p,q)$.
For this choice of $p$ and $q$, 
the energy gap between the third and fourth band is set by the potential strength $2v_2$,
the gap between the second and third band is set by the potential strength $2 v_1$,
and the small gap between the second and third band scales as $\sim v_1 v_2$.
\label{band}}
\end{figure}

\emph{Model --}
We consider the Hamiltonian of a 1D superlattice where one lattice adiabatically slides relative to the other,
\begin{eqnarray}
\label{ham1}
H=\int dx\,\psi^\dagger(x)\left(-\frac{\hbar^2}{2m}\partial_x^2+V_1(x,\varphi)+V_2(x)\right)\psi(x)
\end{eqnarray}
where $\psi(x)$ represents the field operator of the particle, 
$\hbar$ is Planck's constant, and $m$ is the mass of the particle.
The periodic potentials $V_1(x,\varphi)=2v_1\,{\rm cos}(px-\varphi)$ and $V_2(x)=2v_2\,{\rm cos}(qx)$ 
are commensurate, with lattice constants $2\pi/p$ and $2\pi/q$, intensities $v_1$ and $v_2$.
We take the relative phase $\varphi$ to be slowly varying in time.
The period of the Hamiltonian is set by the greatest common divisor 
of $p$ and $q$, i.e., $\kappa\equiv{\rm gcd}(p,q)$, as illustrated in the inset of Fig. \ref{band}.
Treating $1/\kappa$ as the unit length, we redefine
$x\kappa\rightarrow x$, $p/\kappa\rightarrow p$, and $q/\kappa\rightarrow q$.
Treating $E_r=\hbar^2\kappa^2/2m$ as the unit energy,
we redefine $H/E_r\rightarrow H$, $v_1/E_r\rightarrow v_1$, and $v_2/E_r\rightarrow v_2$. 
The dimensionless Hamiltonian in the momentum space is then
\begin{align}
\label{ham2}
H=\sum_k\frac{k^2}{2}\psi_{k}^\dagger\psi_{k}
+\left(v_1e^{-i\varphi }\psi_{k}^\dagger\psi_{k+p}+v_2\psi_{k}^\dagger\psi_{k+q}+ h.c.\right)
\end{align}
Here $\psi_k=\frac{1}{\sqrt{L}}\int dx\, e^{ikx}\psi(x)$, with dimensionless system length $L$ and dimensionless
momentum $k$. Since states of momentum $k$ are coupled only to those of momentum $k+n$ for integer $n$,
we restrict ourselves to the first Brillouin zone ($0\leq k<1$) and rewrite the Hamiltonian,
\begin{align}
\label{ham3}
\notag H=\sum_{0\leq k<1}\sum_{n=-\infty}^\infty\frac{1}{2}(k+n)^2\psi_{n}^\dagger\psi_{n}\\
+\left(v_1e^{-i\varphi}\psi_{n}^\dagger\psi_{n+p}+v_2\psi_{n}^\dagger\psi_{n+q}+h.c.\right)
\end{align}
where we have suppressed the $k$ index, writing $\psi_{n}\equiv\psi_{k+n}$.

To illustrate the resulting band structure, we impose
a cut-off on $n$, and numerically diagonalize the Hamiltonian in Eq. (\ref{ham3}) for $p=2$ and $q=3$.
The lowest four energy bands are shown in Fig. \ref{band},
and even for this simple case the gaps display a range of behaviors for small $v_1$ and $v_2$.
The gap between the third and fourth band is induced by the potential $V_2(x)$, and is proportional to $v_2$ for weak potentials.
The gap between the second and third band is induced by $V_1(x)$, and is proportional to $v_1$.
The small gap between the second and third band is induced
by the combination of these two potentials, which scales as $\sim v_1 v_2$.
In the following section, we will discuss the 
origin of these scalings in the context of understanding the lowest energy gap.

\emph{Band gaps and topology --}
The eigenstates of the Hamiltonian in Eq. (\ref{ham3}) can be found perturbatively in the limit of $v_1,v_2\ll1$.
Suppressing the index $k$, we write $H=H_0+\lambda H_1$, with
\begin{eqnarray}
\label{ham4}
H_0&=&\sum_{n=-\infty}^{\infty}\frac{1}{2}(k+n)^2\psi_{n}^\dagger\psi_{n} \\
\label{ham5}
\lambda H_1&=&\lambda_{p}H_{p}+\lambda_{-p}H_{-p}
+\lambda_{ q}H_{ q}+\lambda_{-q}H_{-q},
\end{eqnarray}
where $H_{p}=\sum_{n=-\infty}^{\infty}\psi_n^\dagger\psi_{n+p}$, $H_{q}=\sum_{n=-\infty}^{\infty}\psi_n^\dagger\psi_{n+q}$,
and $\lambda$ is a formal small parameter, with
$\lambda_{p}=\lambda_{-p}^*=v_1e^{-i \varphi}$ and $\lambda_{q}=\lambda_{-q}^*=v_2$.

For small $\lambda$ and $0\leq k<1$, the eigenstates of the lowest band will be superpositions of $|-1\rangle$ and $|0\rangle$, 
where $|m\rangle=\psi_m^\dagger|{\rm vac}\rangle$. 
We let $\delta k=k-1/2$ denote the distance of $k$ from the band crossing point and assume $\delta k>0$.
The physics for $\delta k<0$ is analogous.
While ordinary perturbation theory works far from the crossing ($\delta k\gg\epsilon$, where $\epsilon$ will be precisely defined below),
one must use higher order degenerate perturbation theory to find the eigenstates for $\delta k\lesssim\epsilon$.
As argued in our supplemental material \cite{supp}, the resulting effective Hamiltonian is of the form
\begin{eqnarray}
H_{\rm eff}=P H_0P+\sum_{\substack{s_+,s_-\geq0\\
r_+,r_-\geq0}}\lambda_{p}^{s_+}\lambda_{-p}^{s_-}
\lambda_{q}^{r_+}\lambda_{-q}^{r_-}\,{\mathcal L}^{(s_+,s_-,r_+,r_-)}
\end{eqnarray}
where $P=|-1\rangle \langle -1|+|0\rangle \langle 0|$, and $s_+,s_-,r_+,r_-$ are integers. The operator
${\mathcal L}^{(s_+,s_-,r_+,r_-)}$ is the contribution to $H_{\rm eff}$
involving the absorption of $\eta=sp+rq$ units of momentum from
the lattices, where $s=s_+-s_-$ and $r=r_+-r_-$. By conservation of momentum, 
$\alpha\equiv\langle-1|{\mathcal L}^{(s_+,s_-,r_+,r_-)}|0\rangle=0$ unless $\eta=1$.
We linearize $H_{\rm eff}$ about $\delta k=0$, and write the operators in the
basis $\left\{|-1\rangle,|0\rangle\right\}$. At the lowest nontrivial order, we have
\begin{eqnarray}
\label{ham6}
{\mathcal H}_{\rm eff}=\left(\begin{array}{cc}
-\frac{1}{2}\delta  k & \alpha\Delta e^{i\chi}\\
\alpha\Delta e^{-i\chi} & \frac{1}{2}\delta k \\
\end{array}
\right)+{\rm const.},
\end{eqnarray}
where $\Delta=v_1^{|r_m|}v_2^{|s_m|},\chi=-s_m \varphi$,
and $s_m,r_m$ correspond to the absolutely smallest solution to the Diophantine equation $sp+rq=1$.
This result agrees with a similar perturbative analysis carried out by
Thouless \emph{et. al.} \cite{TKNN1982} for a related model.

The off-diagonal terms of Eq. (\ref{ham6}) split the energy degeneracy at $\delta k=0$,
and create an energy gap of size $\Delta E_g\equiv2|\alpha\Delta|$. 
For example, if $p=2,q=3$, the absolutely smallest solution to the Diophantine equation has $s_m=-1,r_m=1$,
as $-p+q=1$. Thus the energy gap is $2|\alpha v_1v_2|$, as denoted in Fig. \ref{band}.
For larger $|s_m|$ and $|r_m|$, the energy gap can be extremely small.
Ordinary perturbation theory would have sufficed in the regime where
$\delta k\gg 2|\alpha \Delta|$, allowing us to identify $\epsilon$ as $2|\alpha \Delta|$.
Properties of higher bands can be analyzed similarly.

By analyzing Eq. (\ref{ham6}), we find that the lowest energy eigenstate of Eq. (\ref{ham4}-\ref{ham5}) has the form
\begin{eqnarray}
\label{dstate}
|k,\varphi\rangle=-{\rm sin}\frac{\beta}{2}  e^{i\chi/2}|-1\rangle+{\rm cos}\frac{\beta}{2}  e^{-i\chi/2}|0\rangle+...,
\end{eqnarray}
where ${\rm tan}\beta=-2\alpha\Delta/\delta k$.
The neglected terms are higher order in $v_1$ and $v_2$.
For $\delta k\gg2|\alpha \Delta|$, ${\rm sin}\frac{\beta}{2}\approx1$ and ${\rm cos}\frac{\beta}{2}\approx0$,
and the coefficients are featureless.

Slowly changing $\varphi$ generates an adiabatic current \cite{Thouless1983}.
For a completely filled band, the integrated current in one pumping period ($\varphi$ from $0$ to $2\pi$) is \cite{Niu2010}
\begin{eqnarray}
\label{charge}
\Delta Q=2\pi{\mathcal C}=\int_0^1 dk \int_0^{2\pi}  d\varphi\, \Omega_{k\varphi}
\end{eqnarray}
where the Berry curvature is
\begin{eqnarray}
\Omega_{k\varphi}=
i\left(\partial_{\varphi}\langle k,\varphi|\partial_{k}|k,\varphi\rangle-h.c.\right)=\frac{s_m}{2}\partial_{k}{\rm cos \beta}.
\end{eqnarray}
We see $\Omega_{k\varphi}$ is concentrated near
the location of the energy gap. Integrating the Berry curvature is trivial, yielding
the Chern number ${\mathcal C}=s_m$. 
Although our argument requires that $v_1$ and $v_2$ are small, due to the quantized nature of ${\mathcal C}$,
the result should hold for all nonzero $v_1$ and $v_2$.
In our numerical calculations with larger $v_1,v_2$, we find the curvature is roughly
uniform over the Brillouin zone, but as expected its integral is unchanged.

\begin{figure}[!htb]
\includegraphics[width=8.5cm]{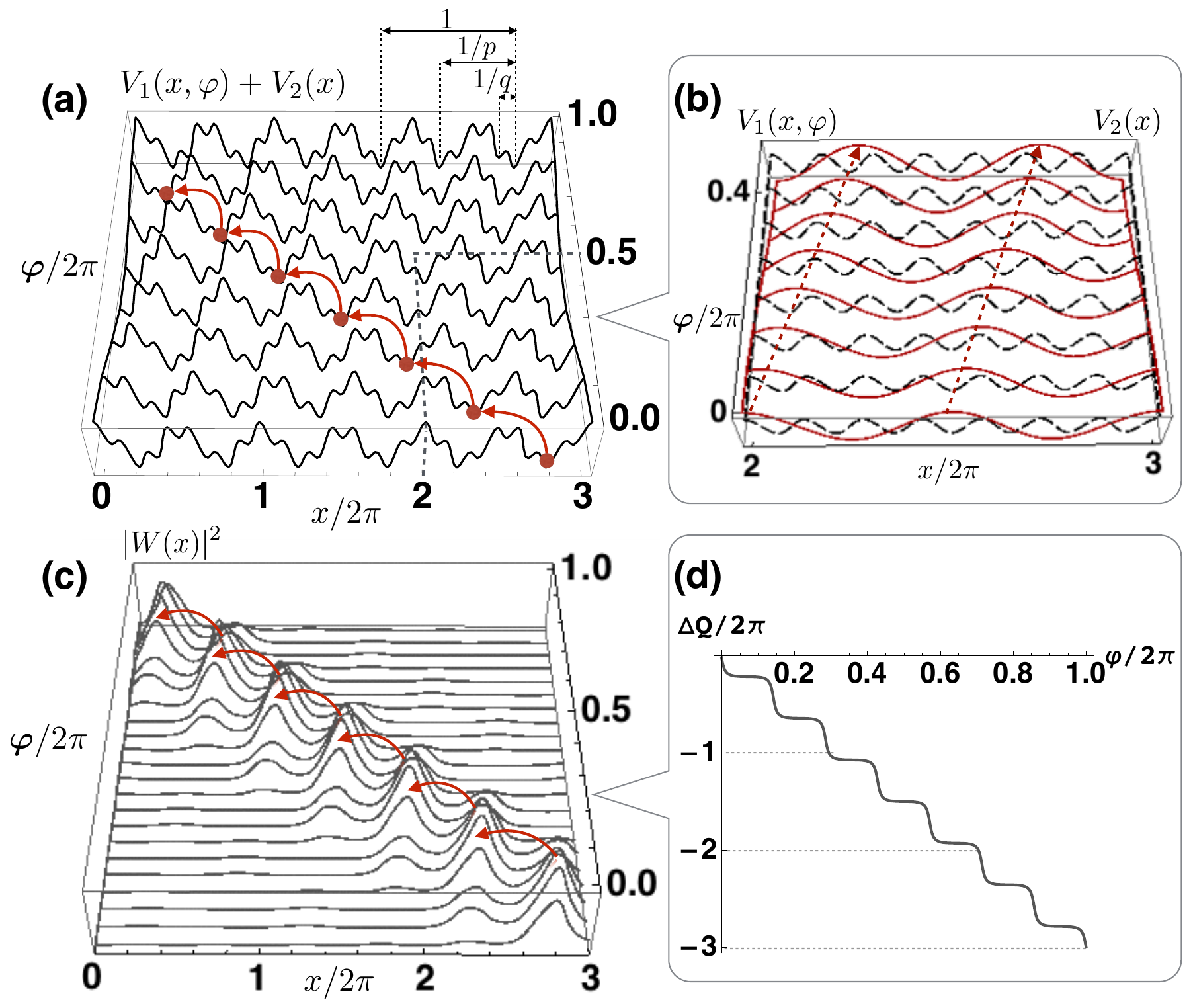}
\caption{(Color online)
(a) Illustration of adiabatic charge transport in a 1D superlattice,
where the particle ``travels" through three unit cells 
to the left when the lattice potential moves to the right by one period. 
Solid lines show the potential $V_1(x,\varphi)+V_2(x)$ for different values of $\varphi$.
Arrows schematically show how the locations of the minima shift discontinuously.
(b) illustrates evolution of two separated potentials of the superlattice:
the right-sliding potential $V_1(x,\varphi)$ (solid red) and the static potential $V_2(x)$ (dashed black).
(c) Evolution of Wannier function. Arrows indicate the ``tunneling" process.
(d) shows the evolution of integrated adiabatic current as a function of $\varphi$.
In these plots we choose $p=2$ and $q=7$, so the Chern number is ${\mathcal C}=s_m=-3$.
Other parameters are $v_1=0.5$ and $v_2=0.25$.
\label{transport}}
\end{figure}

\emph{Anomalous charge pumping --}
By appropriately choosing $p$ and $q$, one can make ${\mathcal C}=s_m$ an arbitrary integer
\cite{Kohmoto1992,Avron2001,Goldman2009,Avron2014}. 
This means that in one pumping cycle a single particle may move arbitrarily far
and/or opposite to the direction of the sliding potential. 
Such long-distance and/or retrograde transport seems unphysical. 
The magic comes from the adiabatic process: If the potential moves sufficiently slowly, the particles always
stay in a global minimum of the potential. Due to the structure of the superlattice, 
a slight motion of the potential could result in a dramatic change of the locations of the global minima (see Fig. \ref{transport}(a)).
Within a small portion of a pumping cycle,
the particles may ``tunnel" to the new global minima which could be a large distance away from the old minima.

To further quantify our interpretations, we calculate the integrated current
\begin{eqnarray}
\label{dcharge}
\Delta Q(\varphi)=\frac{1}{2\pi}\int_{0}^1 dk\int_0^\varphi  d\varphi'\, \Omega_{k\varphi'},
\end{eqnarray}
and the Wannier function at lattice site $j$
\begin{eqnarray}
\label{bloch}
W_j(x,\varphi)=\sum_{0\leq k<1}e^{ik j}\Psi_k(x,\varphi),
\end{eqnarray}
where the Bloch wave function is
\begin{eqnarray}
\label{bloch}
\Psi_k(x,\varphi)=\frac{1}{\sqrt{L}}\sum_{n=-\infty}^{\infty}\langle n|k,\varphi\rangle e^{-i(n+k) x}.
\end{eqnarray}
Here we choose a \emph{smooth} gauge for the Bloch wave function,
so the Wannier function is well localized \cite{Marzari2012}.

Fig. \ref{transport}(d) shows the integrated current as a function of $\varphi$,
calculated from Eq. (\ref{dcharge}) using a similar method to Ref. \cite{Suzuki2005}. We see the function is
``step-like": Flat regions correspond to slow transport, while the particle motion is rapid in the steep regions.
This is further illustrated by the Wannier function in Fig. \ref{transport}(c).
During the slow transport, the Wannier function slowly drifts, while during the rapid transport,
one peak drops in amplitude, and a second peak rises. This corresponds to tunneling.

\begin{figure}[!htb]
\includegraphics[width=8cm]{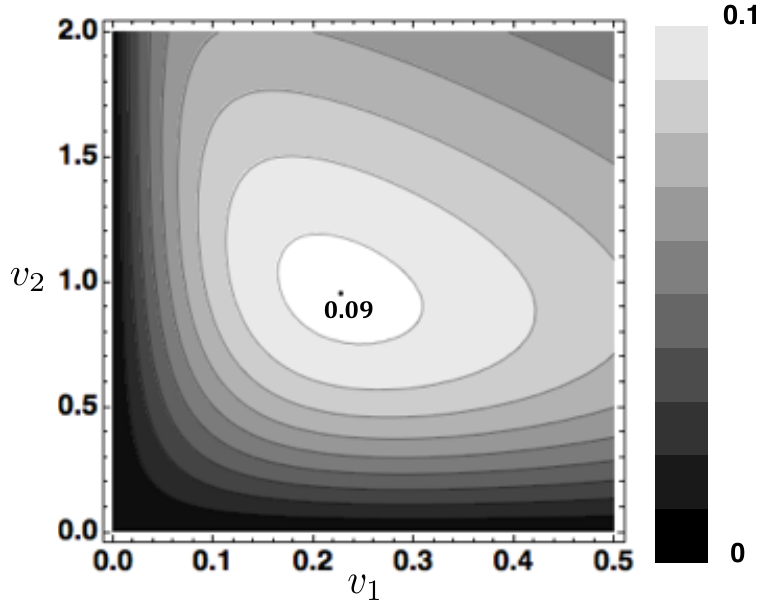}
\caption{Energy gap $\Delta E_g$ as a function of $v_1,v_2$ for $p=2,q=3$.
The gap has a maximium value of $\Delta E_g\approx0.09$ at $v_1=0.23$ and $v_2=0.95$.
\label{gap}}
\end{figure}

For small $v_1,v_2$, the timescale for adiabaticity $\tau$
is related to the size of the gap, $1/\tau\sim|\alpha\Delta|\sim v_1^{|s_m|}v_2^{|r_m|}$.
Thus when the Chern number ${\mathcal C}=s_m$ is large and the potentials
are weak, adiabaticity is hard to maintain in a practical experiment. 
For large $v_1,v_2$, the gap again falls, owing to the large potential barriers. 
Fig. \ref{gap} shows the energy gap $\Delta E_g$ as a function of $v_1$ and $v_2$ for $p=2,q=3$.
The gap has a maximium value of $\Delta E_g\approx0.09$ at $v_1=0.23$ and $v_2=0.95$.
An optimized experiment would be performed with these parameters.

\emph{Experimental proposal --}
To observe this anomalous current, we envision a 
Fermi gas confined to a quasi-1D tube, such that only one transverse mode is occupied.
Although the present analysis is 1D, we expect the phenomena will persist for more general
transverse confinement. Along the tube we engineer 
two longitudinal periodic potentials $V_1(x,\varphi)=2v_1\,{\rm cos}(px-\varphi)$ and $V_2(x)=2v_2\,{\rm cos}(qx)$
via two pairs of counter-propagating laser beams.
The time-dependent phase $\varphi=\delta \omega\, t$ is produced by a frequency difference $\delta \omega$
between two of the beams. To satisfy the adiabatic condition, we require $\hbar\delta \omega\ll\Delta E_g$. 
The resulting adiabatic particle current can be detected by observing the motion of the center of mass of the cloud:
After time $t=2\pi N/\delta\omega$, the center of mass should move a distance $r_c=2\pi {\mathcal C} N/\kappa$.
A dimensionless measure of this displacement is $x_c=\kappa r_c$. The displacement can be measured \emph{in-situ} 
\cite{Chin2009,Greiner2010,Kuhr2010} or after time-of-flight \cite{Bloch2008}.

In modeling this experiment, one must account for the finite cloud size. We include this physics by adding a
harmonic potential along the tube, $V(x)=m\omega_0^2x^2/2$. Such potentials are always found in such experiments.
Within a local density approximation, the lowest band will be filled at the center
of the cloud, but only partially filled near the edge.
Although our Chern number argument only applies to the central region, 
we still expect the center of mass motion to be nearly quantized.
For $\hbar\omega_0\ll v_1,v_2$, and the particle number is much greater than one,
only a very small portion of particles live at boundaries. Our numerical simulations (detailed below) confirm
this results. For a typical experiment, $\omega_0\sim10$ Hz, and $v_1/\hbar, v_2/\hbar\sim100$ kHz \cite{Bloch2002}.

Because of the trap, the displacement $r_c$ cannot be made arbitrarily large. When $m\omega_0r_c^2/2$ is of order of the band gap $\Delta E_g$,
atoms can tunnel to the higher bands. In our numerical simulation, we see that for small $\delta\omega$, the maximum displacement
scales as $1/\omega_0$.
 
\begin{figure}[!htb]
\includegraphics[width=8.5cm]{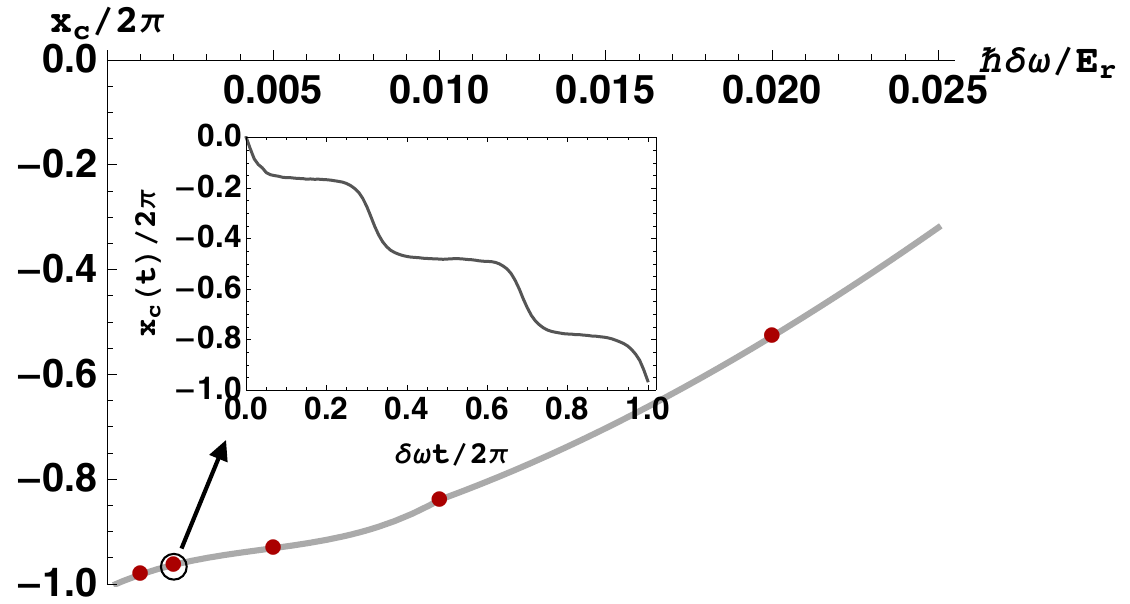}
\caption{(Color online) Displacement of the center of mass (in units of the superlattice)
after one pumping period $T=2\pi/\delta \omega$ for $\nu=63$ fermions in a superlattice with
$p=2,q=3,v_1=0.23 E_r,v_2=0.95 E_r$, and a harmonic trap $\hbar\omega_0=2.2\times 10^{-3}E_r$.
Physically, $\delta \omega$ is the detuning between the beams producing the lattice with wave-number $p$.
We see $x_c/2\pi\rightarrow C=-1$ as $\delta \omega$ decreases.
Inset shows the evolution of the center of mass for $\hbar\delta \omega=0.002E_r$. [c.f. Fig. \ref{transport}(d)]
\label{simu}}
\end{figure}
 
\emph{Numerical simulation --}
In order to see the feasibility of our experimental proposal, 
we numerically simulate the dynamical evolution of a 1D Fermi gas.
We take the many-body state to be a Slater determinant, made up from
single-particle wave functions $\psi_i(x,t)$ with $1\leq i\leq \nu$, where $\nu$ is
the number of fermions. At time $t=0$, $\psi_i(x,0)$ is the $i$th eigenstate
of the Hamiltonian. We evolve $\psi_i(x,t)$ via the time-dependent single-particle Schr\"{o}dinger equation,
and then calculate the center of mass $x_c(t)\equiv1/\nu\sum_{i=1}^{\nu}\int x|\psi_i(x,t)|^2 dx$.
Fig. \ref{simu} shows the results for $p=2,q=3$ where the Chern number is ${\mathcal C}=-1$.
We see $x_c<0$, meaning that the particles travel in the \emph{opposite} direction to the sliding
potential. Remarkably this retrograde motion persists even for relatively large $\delta \omega$.
As $\delta \omega\rightarrow0$ the motion becomes quantized.
A typical experiment has $E_r/\hbar\sim100$ kHz \cite{Bloch2002}, so 
the Chern number ${\mathcal C}=-1$ is readily extracted when $\delta \omega\lesssim 200$ Hz.
The inset of Fig. \ref{simu} shows the evolution of the center of mass in one pumping cycle for $\hbar\delta \omega=0.002E_r$.
We see the function is ``step-like", similar to the ideal case (no harmonic trap and adiabatic) in Fig. \ref{transport}(d).

\emph{Acknowledgement --}
R. Wei thanks Tom\'{a}s Arias for useful discussions on computing the localized Wannier states.
We acknowledge support from ARO-MURI Non-equilibrium Many-body Dynamics grant (W911NF-14-1-0003).

\section{Supplemental material}
Here we derive an effective Hamiltonian
for Eq. (\ref{ham4}-\ref{ham5}). For small $\lambda$
Since the eigenstates of the lowest band will be superpositions of $|-1\rangle$ and $|0\rangle$,
motivating projection operators
\begin{eqnarray}
P&=&|-1\rangle \langle -1|+|0\rangle \langle 0|\\
Q&=&1-P.
\end{eqnarray}
The states $|m\rangle=\psi_m^\dagger|{\rm vac}\rangle$, satisfy $H_0|m\rangle=\frac{1}{2}(k_x+m)^2|m\rangle$.
We seek eigenstates $H|\psi\rangle=E|\psi\rangle$. We break the wave function into two parts 
\begin{eqnarray}
|\psi\rangle=P|\psi\rangle+Q|\psi\rangle\equiv|\psi_0\rangle+|\psi_{\rm ex}\rangle,
\end{eqnarray}
where $|\psi_0\rangle$ is in the low energy sector, and $|\psi_{\rm ex}\rangle$ is a superposition of
the higher-energy states. The eigen-equation is then decoupled into two equations
\begin{eqnarray}
\label{p6}
P H|\psi\rangle&=&PE|\psi\rangle=E|\psi_{0}\rangle\\
\label{p7}
Q H|\psi\rangle&=&QE|\psi\rangle=E|\psi_{\rm ex}\rangle.
\end{eqnarray}
Inserting the identity $P^2+Q^2=P+Q=1$ on the left hand side of Eq. (\ref{p6})-(\ref{p7}) and
substituting $|\psi_{\rm ex}\rangle$ in terms of $|\psi_0\rangle$, we obtain a closed equation for $|\psi_0\rangle$,
\begin{eqnarray}
H_{\rm eff}|\psi_0\rangle=E|\psi_0\rangle,
\end{eqnarray}
where 
\begin{eqnarray}
\label{p5}
H_{\rm eff}\equiv PHP+PHQ\frac{1}{E-Q HQ}QHP.
\end{eqnarray}
Using the identity $PH_0 Q=0$ and expanding the second term of Eq. (\ref{p5}), we obtain
\begin{widetext}
\begin{eqnarray}
H_{\rm eff}=PH_0P+\lambda P H_1P+\lambda^2 P H_1Q\sum_{j=0}^\infty\frac{1}{E-Q H_0 Q}
\left(\lambda \frac{Q H_1 Q}{E-Q H_0 Q}\right)^jQ H_1P.
\end{eqnarray}
\end{widetext}
This equation can be written as
\begin{align}
H_{\rm eff}=P H_0P+\sum_{\substack{s_+,s_-\geq0 \\ r_+,r_-\geq0}}\lambda_{p}^{s_+}\lambda_{-p}^{s_-}
\lambda_{q}^{r_+}\lambda_{-q}^{r_-}\,{\mathcal L}^{(s_+,s_-,r_+,r_-)}
\end{align}
where the momentum conservation implies that 
$\alpha\equiv\langle-1|{\mathcal L}^{(s_+,s_-,r_+,r_-)}|0\rangle=0$ unless $sp+rq=1$, 
where $s=s_+-s_-$ and $r=r_+-r_-$. In our problem, the lowest order contribution to $\alpha$
has either $s_+=0$ or $s_-=0$. Similarly $r_+=0$ or $r_-=0$. 
The lowest order contribution to the diagonal elements of $H_{\rm eff}$ corresponds to an identity matrix.

Linearizing $H_{\rm eff}$ about $\delta k=0$, and writing the operators in the
basis $\left\{|-1\rangle,|0\rangle\right\}$, we have
\begin{eqnarray}
\label{eff}
{\mathcal H}_{\rm eff}=\left(\begin{array}{cc}
-\frac{1}{2}\delta  k & \alpha\Delta e^{i\chi}\\
\alpha\Delta e^{-i\chi} & \frac{1}{2}\delta k \\
\end{array}
\right)+{\rm const.},
\end{eqnarray}
where $\Delta=v_1^{|r_m|}v_2^{|s_m|},\chi=-s_m \varphi$,
and $s_m,r_m$ correspond to the absolutely smallest solution to the Diophantine equation $sp+rq=1$.

\end{document}